\documentclass{article}
\usepackage{spconf,amsmath,graphicx}
\usepackage{pifont}
\usepackage{amsfonts}
\usepackage{multirow}

\usepackage{booktabs}

\title{Cross-Modality and Within-Modality Regularization for Audio-Visual DeepFake Detection}
\name{Heqing Zou, Meng Shen, Yuchen Hu, Chen Chen, Eng Siong Chng, Deepu Rajan}
\address{Nanyang Technological University, Singapore}
\begin{document}
%
\maketitle
\begin{abstract}
Audio-visual deepfake detection scrutinizes manipulations in public video using complementary multimodal cues. Current methods, which train on fused multimodal data for multimodal targets face challenges due to uncertainties and inconsistencies in learned representations caused by independent modality manipulations in deepfake videos. To address this, we propose cross-modality and within-modality regularization to preserve modality distinctions during multimodal representation learning. Our approach includes an audio-visual transformer module for modality correspondence and a cross-modality regularization module to align paired audio-visual signals, preserving modality distinctions. Simultaneously, a within-modality regularization module refines unimodal representations with modality-specific targets to retain modal-specific details. Experimental results on the public audio-visual dataset, FakeAVCeleb, demonstrate the effectiveness and competitiveness of our approach.

\end{abstract}
\begin{keywords}
Audio-visual fusion, deepfake detection, contrastive learning, representation regularization
\end{keywords}

\section{Introduction}
\label{intro00}


Advances in multimedia generation algorithms, including Variational Autoencoders (VAE, \cite{DBLP:journals/corr/KingmaW13}), Generative Adversarial Networks (GAN, \cite{goodfellow2020generative}), and Diffusion models, have enabled the widespread creation and use of synthetic content with minimal expertise. While traditionally applied in film and television production \cite{wu2021f3a}, the proliferation of fabricated media on the internet poses significant public threat \cite{dolhansky2020deepfake}. To combat the impact of deepfake manipulation in daily life, many studies focus on identifying manipulated content, such as face swapping \cite{nirkin2019fsgan} and audio cloning \cite{jia2018transfer}.

\begin{figure}[t]
  \centering
  \includegraphics[width=0.98\linewidth]{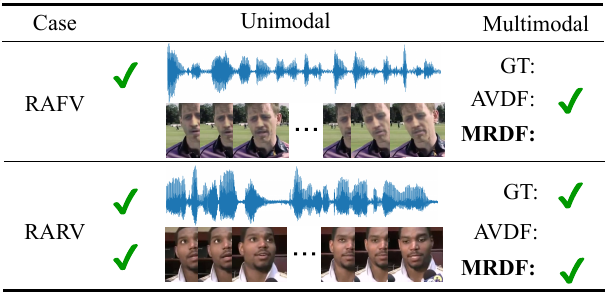}
  \vspace{-0.3cm}
  \caption{\small Proposed \textbf{M}odality-\textbf{R}egularization-based \textbf{D}eep\textbf{F}ake (\textbf{MRDF}) detection on RealAudio-FakeVideo (RAFV) and RealAudio-RealVideo (RARV) categories. (AVDF: Baseline Audio-Visual DeepFake detection, GT: Ground-Truth) }
  \vspace{-0.3cm}
  \label{fig01}
\end{figure}


\begin{figure*}[ht]
\centering
\includegraphics[width=0.69\textwidth]{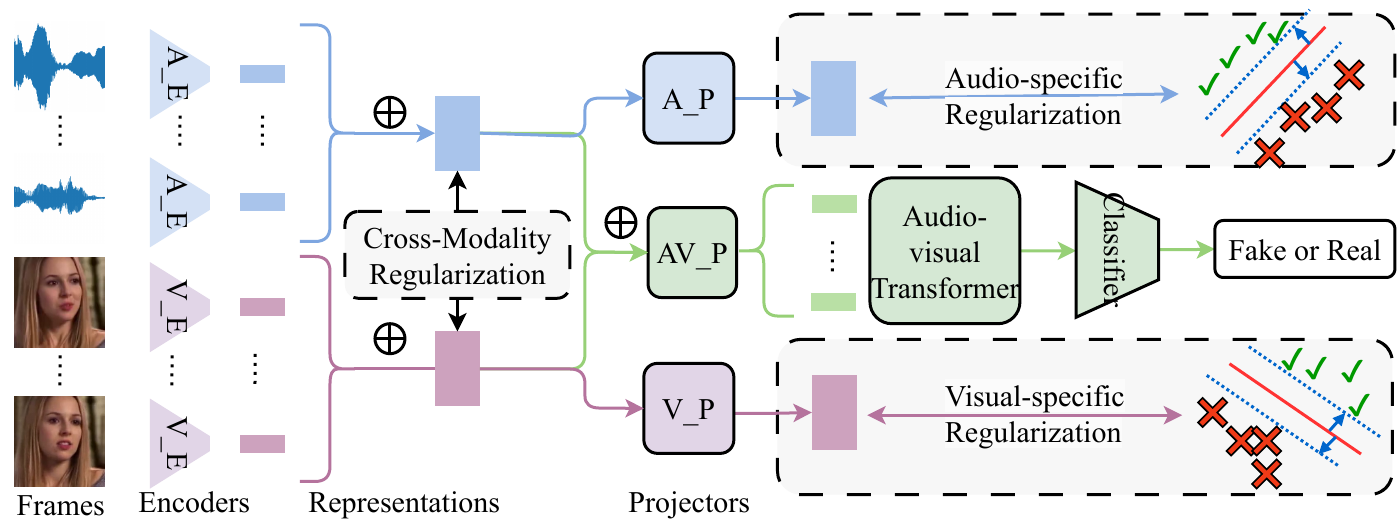}
\vspace{-0.3cm}
\caption{\small Our proposed approach consists of A$\_$E and V$\_$E, representing the audio and video frame encoders. A$\_$P, V$\_$P, and AV$\_$P are the audio feature projector, video feature projector, and audio-visual feature projector, respectively. The symbol $\oplus$ denotes feature concatenation.}
\vspace{-0.3cm}
\label{fig02}
\end{figure*}

%

Unimodal deepfake detection relies on a single modality, such as visual using image-only or video-only signals \cite{hu2022finfer}, or audio using audio signals \cite{yi2022add}. However, these methods are constrained by their input modality and face challenges in addressing real-world scenarios with manipulations spanning multiple modalities. To address this challenge, recent research has focused on multimodal deepfake detection, particularly audio-visual deepfake detection \cite{zhou2021joint, cozzolino2023audio}. These approaches concurrently learn multimodal representations from both audio and visual signals, allowing them to detect deepfakes involving audio or visual manipulations.

Audio-visual deepfake detection can be classified into two categories based on joint learning of audio and visual information. The first method involves training relatively independent modules and making the decision based on the correlation between learned embeddings. Chugh et al. \cite{chugh2020not} employ the Modality Dissonance Score (MDS) to assess dissimilarity between audio and visual modalities in videos, identifying fake videos with high MDS scores. Similarly, techniques like VFD \cite{cheng2022voice} and POI-Forensics \cite{cozzolino2023audio} use contrastive learning to enhance the similarity between audio and visual modalities in authentic videos. However, issues may arise, such as when the relationship between audio and video breaks down when both modalities are manipulated \cite{cheng2022voice}.

In contrast to modality similarity computations, the second approach combines audio-visual representations extracted from unimodal features and maps them towards multimodal objectives. Examples of this approach include Joint-learning \cite{zhou2021joint} and AVoiD-DF \cite{yang2023avoid}. However, in realistic deepfake videos with independent modality manipulations, joint training with fused information may lead to inaccurate mappings of unimodal data to multimodal targets. \textbf{Fig.\ref{fig01}} illustrates that unimodal representations sharing the same label as multimodal targets may exhibit uncertainty because the model relies on multiple modalities for decision-making. Conversely, unimodal representations sharing a different label than multimodal targets may become inconsistent due to backpropagation from opposing multimodal targets. To address this challenge, Multimodaltrace \cite{raza2023multimodaltrace} proposes to utilize the unimodal labels to explore more information and creates a new multi-class multi-label classification. However, the final performance of these methods may still degrade as fused audio-visual representations interfere with each other when mapping to different classes and labels simultaneously.


To address this challenge, we propose a representation-level approach that employs cross-modality and within-modality regularization to preserve the distinct characteristics and disparities of modalities during multimodal representation learning. The cross-modality module emphasizes retaining modality differences and aligning paired videos during fusion. Unlike VFD \cite{cheng2022voice}, which considers both manipulated modalities as negative samples, we treat videos with a single manipulated modality as negatives. This accommodates scenarios where audio and video lack pairing and where one modality remains unchanged. To maintain modality-specific traits, our margin-based within-modality module merges individual modality features with specific targets, possibly derived from final multimodal objectives. Additionally, we enhance audio-visual correspondence through integration with the audio-visual transformer module. Our approach is evaluated on FakeAVCeleb and our code, data processing, and dataset-related resources are readily available \footnote{https://github.com/Vincent-ZHQ/MRDF}.


\section{Methodology}
\label{mtd0}
In this section, we introduce our deepfake detection framework (\textbf{Fig.\ref{fig02}}), comprising unimodal feature extraction and audio-visual fusion modules. Expanding on audio-visual representation learning, we elaborate on our cross-modality and within-modality regularization approach, enhancing the effectiveness of our multimodal representations.

\vspace{-0.15cm}
\subsection{Audio-visual deepfake detection}
\label{mtd1}
The audio and visual channel inputs of the input video are denoted as $x_a$ and $x_v$, respectively. Both inputs are sequential, with $T_a$ and $T_v$ frames. In addition to overall multimodal deepfake detection using only $y_m$ as targets, we include modality-specific information by merging each modality's data with their respective labels $y_a$ and $y_v$. 

\vspace{-0.15cm}
\subsubsection{Feature extraction}
\label{mtd11}
Sequential audio and visual features are extracted using separate frame-level encoders. For audio input $x_a$ and visual input $x_v$ we represent $t^{th}$ frame feature as $f_{a(t)}= \mathcal{F}_{\Phi^{a}}(x_{a(t)})$ and $f_{v(t)}=\mathcal{F}_{\Phi^{v}}(x_{v(t)})$
where $\mathcal{F}{\Phi^{a}}$ and $\mathcal{F}{\Phi^{v}}$ are the modality-specific feature extractors. After concatenation, we obtain the combined frame features $f_{a}$ and $f_{v}$.

\vspace{-0.15cm}
\subsubsection{Audio-visual fusion}
\label{mtd12}
Following AV-Hubert \cite{shi2021learning}, we employ concatenation to fuse the extracted audio and visual features. The projected multimodal features are then passed through the transformer layers:
\begin{equation}
    \setlength{\abovedisplayskip}{2pt}
    \setlength{\belowdisplayskip}{2pt}
    f_{m} = \mathcal{F}_{\Phi^{m}}(\mathcal{P}_{\Phi^{p}}(x_{a} \oplus x_{v})))
\end{equation}
where $f_m$ represents the processed multimodal features. $\mathcal{P}_{\Phi^{p}}$ is the projection layer, and $\mathcal{F}_{\Phi^{m}}$ is the transformer module. 

\vspace{-0.15cm}
\subsection{Regularization}
\label{mtd2}
Common audio-visual deepfake detection methods often result in uncertain and inconsistent modality representations, which hinders the robustness of multimodal detection. To address this, we introduce our cross-modality and within-modality regularization modules.

\vspace{-0.15cm}
\subsubsection{Cross-modality regularization}
\label{mtd21}
In line with other multimodal methods \cite{ zou-etal-2023-unis, DBLP:conf/interspeech/ChenHHZQC22}, we employ contrastive learning to minimize the disparity between paired visual and audio features. Differing from VFD \cite{cheng2022voice}, we consider samples with one or more manipulated modalities as negative pairs, as these features are inevitably unpaired. The cross-modality regularization $\mathcal{L}_\text{cmr}$ across the total $N$ samples can be defined as follows:
\begin{equation}
    \setlength{\abovedisplayskip}{2pt}
    \setlength{\belowdisplayskip}{2pt}
    \sum_{i=1}^{N}[ y_c^i \ast {(1 - d(x_a^{i}, x_v^{i}))} + (1-y_c^i) \ast \text{max}(0, d(x_a^i, x_v^i)))]
\end{equation}
where $y_c^i$ represents the label for sample $i$ in cross-modal regularization, taking 1 for paired audio-visual samples and 0 for others. The term $d(x_a^i, x_v^i) = \frac{x_a \cdot x_v}{\lVert x_a \rVert \lVert x_v \rVert}$ calculates the cosine similarity between the extracted audio and visual features.

\vspace{-0.15cm}
\subsubsection{Within-modality regularization}
\label{mtd22}
To maintain the integrity of individual modality features, we align unimodal representations with their respective targets separately using within-modality regularization loss $\mathcal{L}_{\text{wmr}}$, which including two branches, visual-specific regularization and audio-specific regularization. We propose two modality-specific regularization methods for analysis. The margin-based regularization for modality $n$ is given by
\begin{equation}
    \setlength{\abovedisplayskip}{2pt}
    \setlength{\belowdisplayskip}{2pt}
     \mathcal{L}_\text{wmr-margin}^{n} = \sum_{i=1}^{N} [ \sum_{y_n^i=y_n^j} ({1 - d_n^{ij}))} + \sum_{y_n^i \ne y_n^j} \text{max}(0,  d_n^{ij}) - \alpha_{n}))]
\end{equation}
where $d_n^{ij}$ represents the cosine similarity between target samples $i$ and $j$ of modality $n \in {a, v}$ features, with $\alpha_n$ as the margin value for modality $n$. The cross-entropy-based modality-specific regularization for modality $n$ is as follows:
\begin{equation}
    \setlength{\abovedisplayskip}{3pt}
    \setlength{\belowdisplayskip}{3pt}
      \mathcal{L}_\text{wmr-ce}^{n}=\sum_{c=1}^{k} (y_m^{c} \ast \text{log} \frac{\text{exp}(f_{\theta}^{n}(x_n)^c)}{\sum_{c=1}^{k} \text{exp}(f_{\theta}^{n}(x_n)^c)})
\end{equation}
where $k=2$ for binary deepfake detection and $f_{\theta}^{n}(\cdot)$ is the unimodal classifier module for modality $n$.

\subsection{Learning objective}
\label{mtd3}
We use the following cross-entropy loss $\mathcal{L}_\text{ce}$ to detect the audio-visual deepfakes:
\begin{equation}
    \setlength{\abovedisplayskip}{3pt}
    \setlength{\belowdisplayskip}{3pt}
     \mathcal{L}_\text{ce} = -\sum_{c=1}^{k} (y_m^{c} \ast \text{log} \frac{\text{exp}(f_{\theta}^{m}(x_a, x_v)^c)}{\sum_{c=1}^{k} \text{exp}(f_{\theta}^{m}(x_a, x_v)^c)})
\end{equation}
where $k=2$ for binary deepfake detection and $f_{\theta}^{m}(\cdot)$ is the multimodal classifier module. The total loss to optimize the proposed audio-visual deepfake detection method with modality-independent and modality-specific regularization is:
\begin{equation}
    \setlength{\abovedisplayskip}{2pt}
    \setlength{\belowdisplayskip}{2pt}
     \mathcal{L}_\text{avdf} = \lambda_\text{ce} \mathcal{L}_\text{ce} + \lambda_\text{cmr} \mathcal{L}_\text{cmr} + \lambda_\text{wmr} \mathcal{L}_\text{wmr}
\end{equation}
where $\lambda_\text{ce}$, $\lambda_\text{cmr}$ and $\lambda_\text{wmr}$ are the weights for each loss.

\section{Experiment}
\label{exp0}

\subsection{Datasets}
\label{exp1}
We evaluate our method on the public audio-visual deepfake detection datasets: FakeAVCeleb \cite{khalid2021fakeavceleb}. FakeAVCeleb consists of 500 real videos and over 20,000 fake videos, spanning five ethnic groups, each with 100 real videos from 100 subjects. For equitable comparisons, we employ a balanced setting with a 1:1:1:1 ratio across four categories,  FakeAudio-FakeVideo (FAFV), FakeAudio-RealVideo (FARV), RealAudio-FakeVideo (RAFV), and RealAudio-RealVideo (RARV), and utilize a 5-fold-cross-validation strategy.

\subsection{Experimental setup}
\label{exp2}
Following prior audio-visual methods \cite{ma2021end, shi2021learning}, we employ a linear projection layer and modify a ResNet-18 for the audio and visual encoders. The audio-visual transformer module comprises 12 transformer blocks. Our model is trained for 30 epochs using Adam optimization, with an initial learning rate of $1e-3$ and a batch size of 64. We assign equal weights to the sub-losses for balanced optimization. The margin values for both audio and visual modalities are empirically set to 0.


\begin{table}
\centering
\setlength\tabcolsep{12pt}
\vspace{-0.2cm}
\caption{\small Performances comparison of audio-visual deepfake detection with SOTA methods on FakeAVCeleb.}
\vspace{0.15cm}
\resizebox{0.47\textwidth}{!}{
\begin{tabular}{p{12em}|cccc}
\toprule
     {Method} &  & {ACC $\uparrow$} & {AUC $\uparrow$}   \\
    \midrule
    VFD \cite{cheng2022voice}  & & 81.52 & 86.11  \\
    AVOiD-DF \cite{yang2023avoid} & &  83.7 & 89.2  \\
    DST-Net \cite{ilyas2023avfakenet} &  & 92.59 & - \\
    Ensemble \cite{hashmi2022multimodal}  & & 89 & -   \\
    Multimodaltrace \cite{raza2023multimodaltrace}  & & 92.9 & -   \\
    \midrule
    MRDF-Margin  &  & 93.40 & 91.80  \\
    MRDF-CE  &  & 94.05  & 92.43  \\
\bottomrule
\end{tabular}}
\label{tab01}
\vspace{-0.4cm}
\end{table}

\begin{table}
\centering
\setlength\tabcolsep{6pt}
\caption{\small Performances comparison with different uni-modality-constraint methods on FakeAVCeleb of 5-fold cross-validation.}
\vspace{0.15cm}
\resizebox{0.47\textwidth}{!}{
\begin{tabular}{p{10em}cccc}
\toprule
     {Model} & {Method}  & {ACC $\uparrow$} & {AUC $\uparrow$} &   \\
     \midrule
     Multimodal AVDF \cite{khalid2021evaluation} & Mixing & 89.05 & 88.30  \\
     Ensemble AVDF  \cite{hashmi2022multimodal} & Ensemble & 91.15 & 89.90 \\
     Multimodaltrace \cite{raza2023multimodaltrace}  & Multi-label  & 92.25 & 89.83  \\
     Multimodaltrace \cite{raza2023multimodaltrace} & Multi-class  & 92.60 & 90.93  \\
     \midrule
     MRDF-Margin  & Regularization &  93.40 & 91.80   \\
     MRDF-CE  & Regularization &  94.05  & 92.43 \\
\bottomrule
\end{tabular}}
\label{tab02}
\vspace{-0.3cm}
\end{table}

\begin{table*}
\centering
\setlength\tabcolsep{5pt}
\vspace{-0.3cm}
\caption{\small Ablation study of the proposed modality-regularization-based method on FakeAVCeleb.}
\vspace{0.1cm}
\resizebox{0.95\textwidth}{!}{
\begin{tabular}{p{8em}lcccccccccccc}
\toprule
     \multirow{2}{*}{Model} & \multirow{2}{*}{Regularization} &  & \multicolumn{3}{c}{Real}   &  & \multicolumn{3}{c}{Fake} & & \multicolumn{2}{c}{All} \\
     \cmidrule{4-6}  \cmidrule{8-10} \cmidrule{12-13}
     & & & {Precision} & {Recall} & {F1-Score} & & {Precision} & {Recall} & {F1-Score} & & {Accuracy} & {AUC} \\
     \midrule
     Multimodal AVDF & N.A. & & 74.59 & 86.80 & 80.09 &  & 95.30 & 89.80 & 92.44 & & 89.05 & 88.30  \\
     &  $w.$ $\mathcal{L}_\text{cmr}$ & & 82.99 & 88.40 & 85.59 &  & 96.05 & 93.93 & 94.98 & & 92.55 & 91.17 \\
     &  $ w.$ $\mathcal{L}_\text{wmr-margin}$  & & 83.18 & 89.20 & 85.10 &  & 96.31 & 93.93 & 95.10 & & 92.75 & 91.57 \\
    {MRDF-Margin}  & $ w.$ $\mathcal{L}_\text{cmr}, \mathcal{L}_\text{wmr-margin}$ & & 85.69 & 88.60 & 87.05 &  & 96.16 & 95.00 & 95.57 & & 93.40 & 91.80 \\
     & $ w.$ $\mathcal{L}_\text{wmr-ce}$  & & 88.30 & 89.00 & 88.64 &  & 96.33 & 96.07 & 96.20 & & 94.30 & 92.53 \\
    {MRDF-CE} & $ w.$ $\mathcal{L}_\text{cmr}, \mathcal{L}_\text{wmr-ce}$ & & 87.46 & 89.20 & 88.22 &  & 96.40 & 95.76 & 96.02 & & 94.05 & 92.43 \\
\bottomrule
\end{tabular}}
\label{tab03}
\vspace{-0.4cm}
\end{table*}

\begin{figure}[t]
  \centering
  \includegraphics[width=0.85\linewidth]{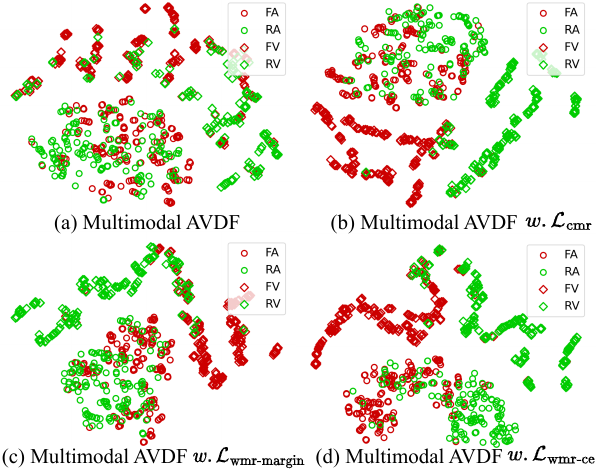}
  \vspace{-0.3cm}
  \caption{\small T-SNE visualization of the audio and visual representations before fusion of the ablation study methods.}
  \vspace{-0.5cm}
  \label{fig05}
\end{figure}

\section{Result and analysis}
\label{ra0}
In this section, we evaluate our regularization-based method's performance and conduct an ablation study to provide a detailed analysis of our proposed method.

\subsection{Main results and comparison}
\label{ra1}
In \textbf{Table \ref{tab01}}, our MRDF achieves the highest performance, with a classification accuracy of $94.05\%$ and an AUC score of $92.43\%$ using identity-independent five-fold cross-validation on FakeAVCeleb. Comparing our method to other unimodal-deepfake-constrained methods on the same 5-fold cross-validation validation strategy (\textbf{Table \ref{tab02}}), all constraint-based methods outperform multimodal audio-visual deepfake detection (AVDF, \cite{khalid2021evaluation}). This suggests that both methods, aligning mixed modality representations with new classification heads and our representation-regularization-based approach, contribute to reliable multimodal deepfake detection. Moreover, our model exhibits superior performance with two different regularization methods, demonstrating efficiency without the need for introducing a new classification head.


\vspace{-0.8cm}

\subsection{Ablation study}
\label{ra2}

We apply representation-level cross-modality and within-modality regularization to enhance audio-visual deepfake detection. Cross-modality regularization differentiates between paired and unpaired modalities, bolstering confidence in real samples through cross-modal alignment. Within-modality regularization amalgamates unimodal label information, enhancing modality distinguishability. \textbf{Table \ref{tab03}} exhibits improved performance with both introduced cross-modality and within-modality regularization methods. However, cross-modality regularization exerts a milder influence on the cross-entropy-based method than the margin-based approach, as binary cross-entropy loss can align cross-modal representations, as seen in CLIP \cite{radford2021learning}. In \textbf{Fig.\ref{fig05}}, we visually compare unimodal representations of audio and visual modalities before fusion. The proposed cross-modality regularization refines the alignment of paired audio-visual representations (green) in (b) compared to the baseline method in (a). Within-modality regularization heightens the distinguishability of unimodal representations with different targets, as depicted in (c) compared to (a). Finally, cross-entropy-based regularization renders clearer distinctions among different classes.

\subsection{Analysis and visualization}
\label{ra3}
\begin{figure}[t]
  \centering
  \includegraphics[width=0.95\linewidth]{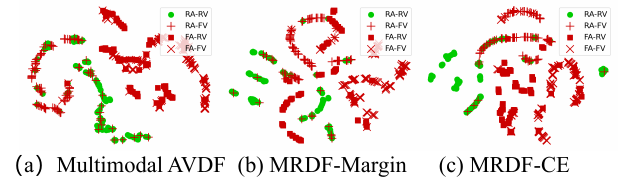}
  \vspace{-0.2cm}
  \caption{\small T-SNE visualization of the deepfake prediction of (a) Multimodal AVDF and our proposed (b) MRDF-Margin (c) MRDF-CE.}
  \vspace{-0.5cm}
  \label{fig04}
\end{figure}

\textbf{Table \ref{tab04}} displays classification results for various deepfake scenarios. Both the AVDF baseline and our regularization-based approach correctly identify most fake videos with fake audio. However, our method significantly outperforms the baseline, with lower misclassification rates ($14.0\%$ and $12.6\%$ compared to $30.0\%$) for fake videos with genuine audio. Our regularization constrains the real audio modality and improves multimodal representations with visual fake information for deepfake detection. Additionally, the baseline model misclassifies roughly $13.2\%$ of real samples as fake due to inconsistent unimodal representations of the corresponding genuine modality, leading to erroneous decisions. This issue is mitigated by training the model with constrained unimodal representation learning. We visualize deepfake predictions in \textbf{Fig.\ref{fig04}} using the t-SNE method, highlighting how fake videos with a single fake modality often blend with other categories, particularly those with fake audio. Furthermore, some real videos resemble fake videos when influenced by samples with genuine audio. Our proposed method effectively addresses these misrepresentations, resulting in improved model performance.

\begin{table}
\centering
\setlength\tabcolsep{6pt}
\vspace{-0.4cm}
\caption{\small Classification results for different deepfake scenarios.}
\vspace{0.1cm}
\resizebox{0.43\textwidth}{!}{
\begin{tabular}{p{8em}cccccccccccc}
\toprule
     {Model} &  & {FAFV}  & {FARV}  & {RAFV} & {RARV}  \\
     \midrule
     Multimodal AVDF  &  & 99.8 & 99.6 & 70.0 & 86.8  \\
     MRDF-Margin & & 100.0 & 99.2 & 86.0 & 89.4   \\
     MRDF-CE & & 100.0 & 99.6  & 87.4 & 89.2  \\
\bottomrule
\end{tabular}}
\label{tab04}
\vspace{-0.2cm}
\end{table}

\section{Conclusion}
\label{con0}
This paper presents a representation-level approach to enhance representation learning in audio-visual deepfake detection, addressing uncertainty and inconsistency. We introduce cross-modality and within-modality regularization to improve audio-visual deepfake representation learning, bolstered by an audio-visual transformer module for improved correspondence. Our method demonstrates competitive performance on a public dataset compared to state-of-the-art methods.

\vfill\pagebreak



\small
\bibliographystyle{IEEEtran}
\bibliography{strings,refs}

\end{document}